\documentclass[12pt,showpacs,preprintnumbers,amsmath,amssymb]{revtex4-1}

\usepackage{graphicx}

\begin{document}

\title{The Sznajd model with limited persuasion: competition between high-reputation and hesitant agents}

\author{Nuno Crokidakis $^{1,2}$}
\email{nuno@if.uff.br}

\author{Paulo Murilo Castro de Oliveira $^{1,2}$}
\email{pmco@if.uff.br}
\address{
$^{1}$Instituto de F\'{\i}sica - Universidade Federal Fluminense \\
Av. Litor\^anea s/n \\
24210-340 \hspace{5mm} Niter\'oi - Rio de Janeiro \hspace{5mm} Brazil \\
$^{2}$National Institute of Science and Technology for Complex Systems, Brazil}

\date{\today}

\begin{abstract}

In this work we study a modified version of the two-dimensional Sznajd sociophysics model. In particular, we consider the effects of agents' reputations in the persuasion rules. In other words, a high-reputation group with a commom opinion may convince their neighbors with probability $p$, which induces an increase of the group's reputation. On the other hand, there is always a probability $q=1-p$ of the neighbors to keep their opinions, which induces a decrease of the group's reputation. These rules describe a competition between groups with high reputation and hesitant agents, which makes the full-consensus states (with all spins pointing in one direction) more difficult to be reached. As consequences, the usual phase transition does not occur for $p<p_{c} \sim 0.69$ and the system presents realistic democracy-like situations, where the majority of spins are aligned in a certain direction, for a wide range of parameters.

\end{abstract}


\maketitle

\section{Introduction}

\qquad Social dynamics have been studied through statistical physics techniques in the last twenty years. Among the studied problems, we can cite models of cultural \cite{axelrod}, language \cite{baronchelli} and opinion dynamics \cite{holley,galam,sznajd_model} (for a recent review, see \cite{loreto_rmp}). These models are interesting for physicists because they present order-disorder transitions, scaling and universality, among other typical features of physical systems \cite{loreto_rmp}.

Ising-type models, which can represent systems with binary variables $0$ and $1$ (or $\pm 1$), have been used by physicists in many different areas, such as sociology, politics, marketing and finance \cite{pmco_book, loreto_rmp, stauffer_review, sznajd_review, adriano}. In 2000, an agent-based model was successfully applied to the dynamics of a social system \cite{sznajd_model}. In this model, each agent carries one of two possible opinions, which may be represented by Ising spins. The aim of the Sznajd Model (SM) is the emergence of social collective (macroscopic) behavior due to the interactions among individuals, constituting the microscopic level of a social system \cite{loreto_rmp}.

The original SM consists of a chain of sites with periodic boundary conditions where each site (individual opinion) could have two possible states (opinions) represented in the model by Ising spins (``yes'' or ``no''). A pair of parallel spins on sites $i$ and $i+1$ forces its two neighbors, $i-1$ and $i+2$, to have the same orientation (opinion), while for an antiparallel pair $(i,i+1)$, the left neighbor ($i-1$) takes the opinion of the spin $i+1$ and the right neighbor ($i+2$) takes the opinion of the spin $i$. In this first formulation of the SM two types of steady states are always reached: complete consensus (ferromagnetic state) or stalemate (anti-ferromagnetic state), in which every site has an opinion that is different from the opinion of its neighbors. However, the transient displays an interesting behavior, as pointed by Stauffer et al. \cite{adriano}. Defining the model on a square lattice, the authors in \cite{adriano} considered not a pair of neighbors, but a $2\times 2$ plaquette with four neighbors. Considering that each plaquette with all spins parallel can convince all their eight neighbors, a phase transition was found for an initial density of up spins $d=1/2$. 

A more realistic situation is to consider a probability of persuasion of each site. The Sznajd model is robust against this situation: if one convinces the neighbors only with some probability $p$, and leaves them unchanged with probability $1-p$, still a consensus is reached after a long time \cite{stauffer_review}. Models that consider many different opinions (using Potts' spins, for example) or defined on small-world networks were studied in order to represent better approximations of real communities' behavior (see \cite{stauffer_review} and references therein). In another work, in order to avoid full consensus in the system and makes the model more realistic, Schneider introduced opportunists and persons in opposition, that are not convinced by their neighbors \cite{schneider}, whereas Stauffer considered that only individuals with similar opinions can be convinced \cite{stauffer_2}. In a recent work, another formulation of the Sznajd model with limited persuasion was studied, considering that only high-reputation groups can convince neighbors \cite{sznajd_meu}. 

In this work we analyze a modification of the Sznajd model with reputation studied in \cite{sznajd_meu}. We consider that the reputations may increase and decrease, depending if agents are persuaded or not by high-reputation groups. This fact makes the models' rules intrinsically dynamic, a common feature of evolutionary systems \cite{goldenfeld}. The phase transitions as well as the relaxation times of the model are studied, and the relations between the quantities of interest and the parameters of the model are discussed.


\section{Model}

\qquad The model considered in this paper is a modification of a recente work \cite{sznajd_meu}. Thus, we have considered the Sznajd model \cite{adriano} defined on a square lattice with linear size $L$ and periodic boundary conditions, where an integer number ($R$) labels each player and represents its reputation across the community. The reputation is introduced as a score which is time dependent. The agents start with a random distribution of the $R$ values, and during the time evolution, the reputation of each agent changes according to its capacity of persuasion, following the model's rules. One time step in our model is defined by the following rules:

\begin{enumerate}
\item We randomly choose a 2 $\times$ 2 plaquette of four neighbors.
\item If not all four spins are parallel, leaves its eight neighbors unchanged.
\item If all four center spins are parallel, we calculate the average reputation of the plaquette:
\begin{eqnarray} \nonumber
\bar{R} = \frac{1}{4}\sum_{i=1}^{4}R_{i}~,
\end{eqnarray}
where each term $R_{i}$ represents the reputation of one of the plaquettes' agents \cite{foot1}.
\item Then, we compare the average reputation $\bar{R}$ with the reputations of each one of the 8 sites neighboring the plaquette. Thus, with probability $p$ we follow the rule presented in \cite{sznajd_meu}, i.e., given a plaquette neighbor $j$ with reputation $R_{j}$, this individual $j$ will follow the plaquette opinion if $\bar{R}>R_{j}$. In this case, the reputation of each agent in the 2 $\times$ 2 plaquette is increased by 1 (in other words, the average reputation of the plaquette is increased by 1).
\item On the other hand, with probability $q=1-p$ the neighbor $j$ keeps his opinion, even if we have $\bar{R}>R_{j}$. In this case, the reputation of each agent in the 2 $\times$ 2 plaquette is decreased by 1.
\item We repeat steps $4$ and $5$ for each one of the eight plaquette neighbors.
\end{enumerate}

Notice that agents belonging to a given plaquette may be rewarded or penalized in terms of reputation. A similar approach was analyzed in the Naming game, where a given agent's reputation is increased (decreased) depending on the outcome, i.e., success (failure) of his communication attempt with some other agent \cite{edgardo}. Agent-based models of animal behavior \cite{bonabeau,bonabeau2} and human relations \cite{stauffer_bonabeau} have also considered reputation-like mechanisms in order to take into account different strengths or convincing powers.

The above-mentioned rules define a competition between groups of high-reputation agents and hesitant individuals, that have a probability $q=1-p$ of keeping their opinions despite the influence of plaquettes with high reputation. Due to above microscopic rules, one can expect difficulties to the system to reach the full-consensus steady state, with all spins pointing in a certain direction. Observe that for $p=1$ we recover the model studied in \cite{sznajd_meu}, where democracy-like situations, with the majority of spins pointing in one direction, occur spontaneously, due to the dynamic rules considered in the model. Thus, concerning these realistic situations, we can expect similar results in our case, at least for values of the probability $p$ near unity.



\section{Numerical Results}

\begin{figure}[t]
\begin{center}
\vspace{1.0cm}
\includegraphics[width=0.4\textwidth,angle=0]{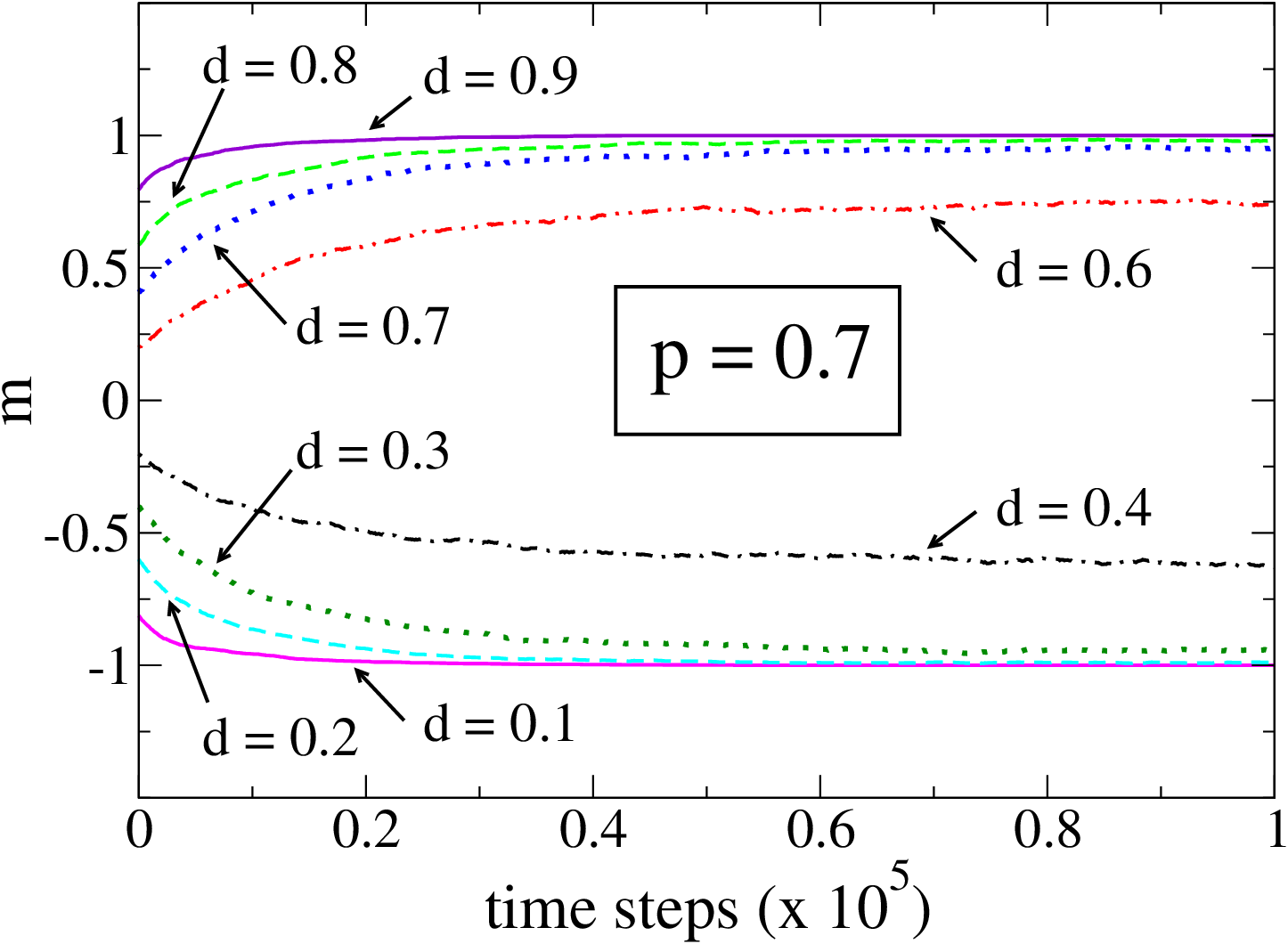}
\hspace{0.5cm}
\includegraphics[width=0.4\textwidth,angle=0]{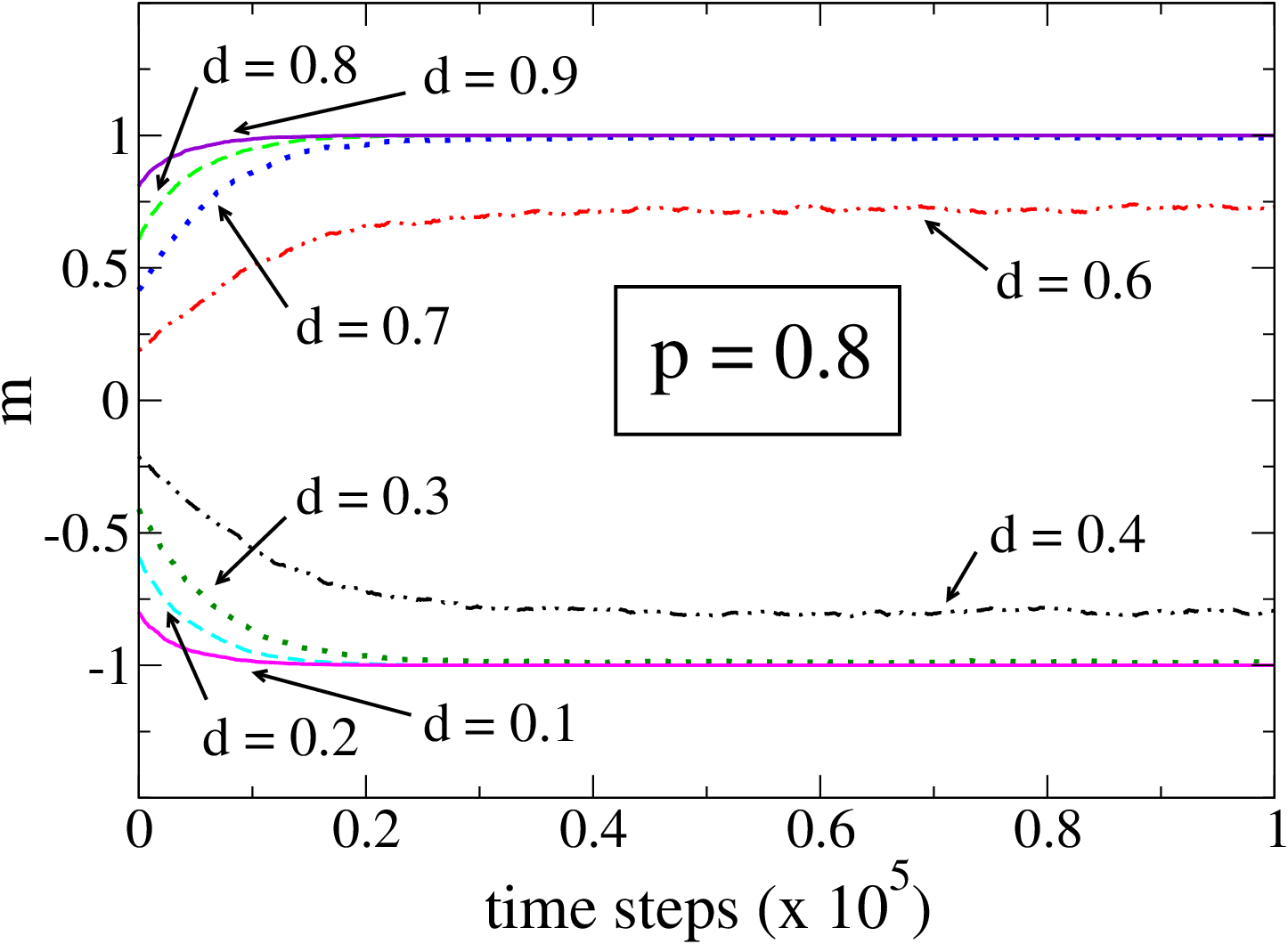}
\end{center}
\caption{Time evolution of the magnetization per spin for $L=73$, some values of the initial density of up spins $d$ and two different values of the probability $p$, namely $p=0.7$ (left side) and $p=0.8$ (right side). We can observe steady states with $|m|<1$, where there is no consensus in the population, as well as full-consensus states, where $|m|=1$.}
\label{fig1}
\end{figure}

\qquad In the simulations, the initial values of the agents' reputation follow a gaussian distribution centered at $0$ with standard deviation $\sigma=5$ \cite{foot2}. We can start studying the time evolution of the ``magnetization per spin", 
\begin{equation}
m=\frac{1}{N}\sum_{i=1}^{N}s_{i} ~,
\end{equation}
\noindent
where $s_{i}=\pm 1$ and $N=L^{2}$ is the total number of agents on the lattice. We exhibit in Fig. \ref{fig1} the magnetization per spin as a function of the number of time steps for $p=0.7$ (left side) and $p=0.8$ (right side) and some values of the initial density of up spins $d$. Observe that the time needed for the system to reach the steady states increases for decreasing values of $p$. This result is easy to understand: when we decrease the value of the probability $p$, the probability $q=1-p$ of the agents to keep their opinions increases, which makes stronger the competition between increasing and decreasing reputations. Observe that for $p<1$ situations with $|m|<1$ emerge spontaneously (see Fig. \ref{fig1}), as in the $p=1$ case \cite{sznajd_meu}. On the other hand, full-consensus states with all spins pointing up or down are also possible in our model, as we can also see in Fig. \ref{fig1}. Thus, the dynamics of competition between reputations leads to the emergence of democracy-like situations in our model. In addition, the results of Fig. \ref{fig1} suggest that for $p<1$ there are transitions between a phase where the full-consensus states with $m=1$ are always reached and another phase where this consensus never occurs. We will analyze this phase transition in the following.

\begin{figure}[t]
\begin{center}
\vspace{1.0cm}
\includegraphics[width=0.5\textwidth,angle=0]{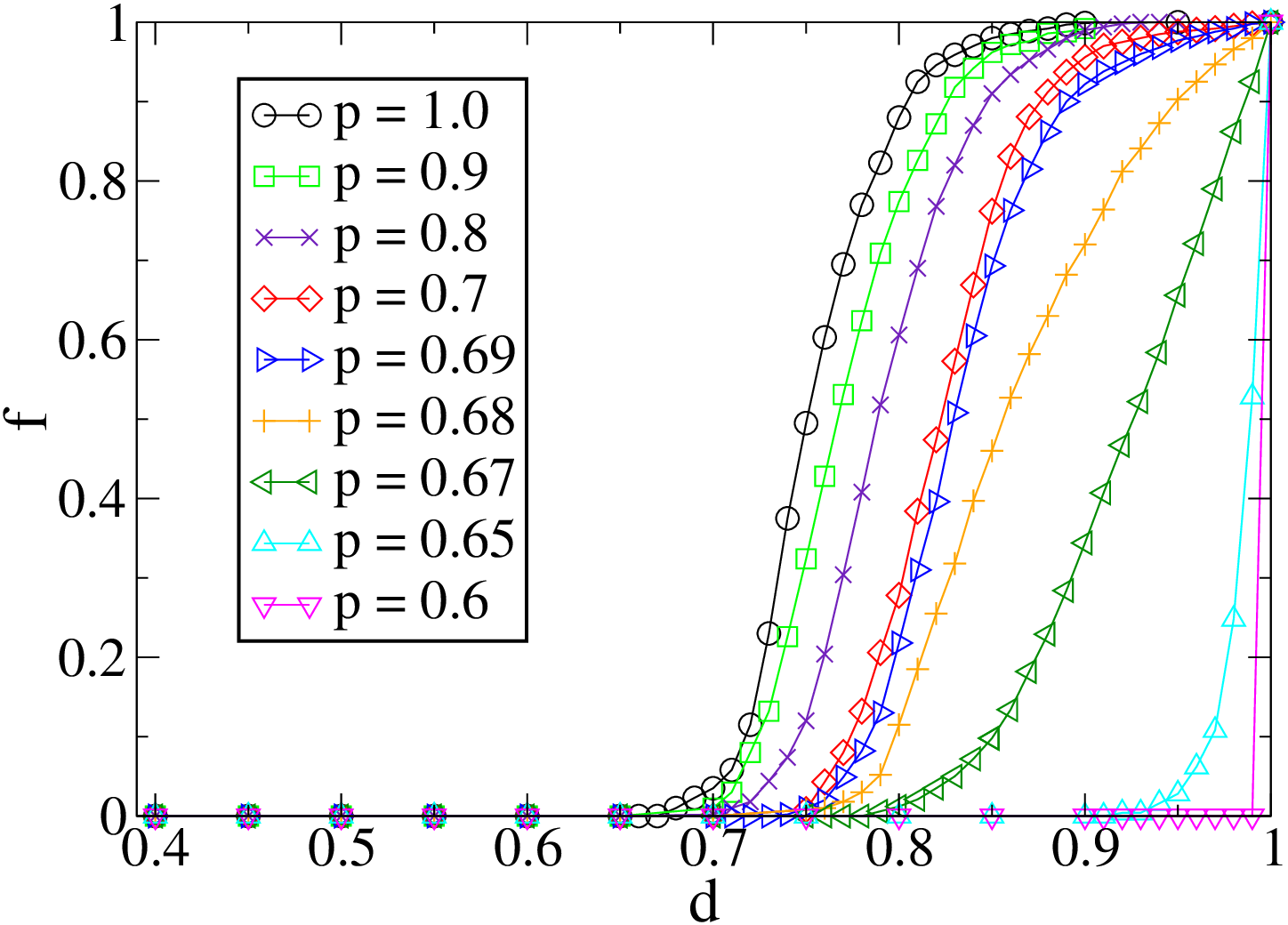}
\end{center}
\caption{Fraction $f$ of samples which show all spins up when the initial density of up spins $d$ is varied in the range $0.4\leq d\leq 1.0$, for $L=101$ and typical values of the probability $p$. A phase transition can be already observed to occur at some critical point $p_{c}$ ($0.68< p_{c} < 0.69$), better analysed in the sequence.}
\label{fig3}
\end{figure}

\begin{figure}[t]
\begin{center}
\vspace{1.0cm}
\includegraphics[width=0.5\textwidth,angle=0]{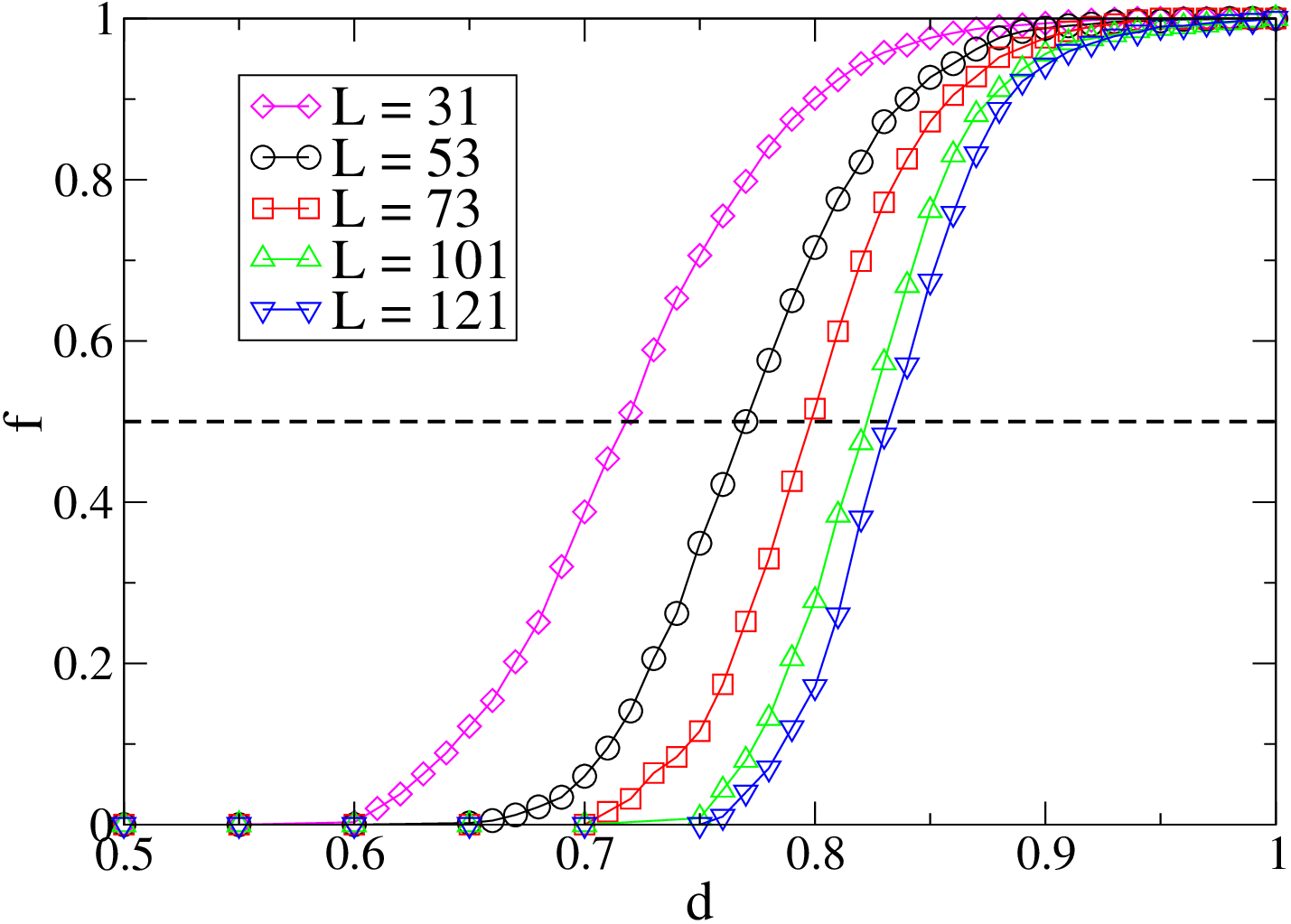}
\end{center}
\caption{Fraction $f$ of samples which show all spins up when the initial density of up spins $d$ is varied in the range $0.5\leq d\leq 1.0$, for $p=0.7$ and some values of the lattice size $L$. The dashed line shows that the value of $d$ corresponding to the level $f=0.5$ increases for increasing values of $L$.}
\label{fig4}
\end{figure}

\begin{figure}[t]
\begin{center}
\vspace{1.0cm}
\includegraphics[width=0.46\textwidth,angle=0]{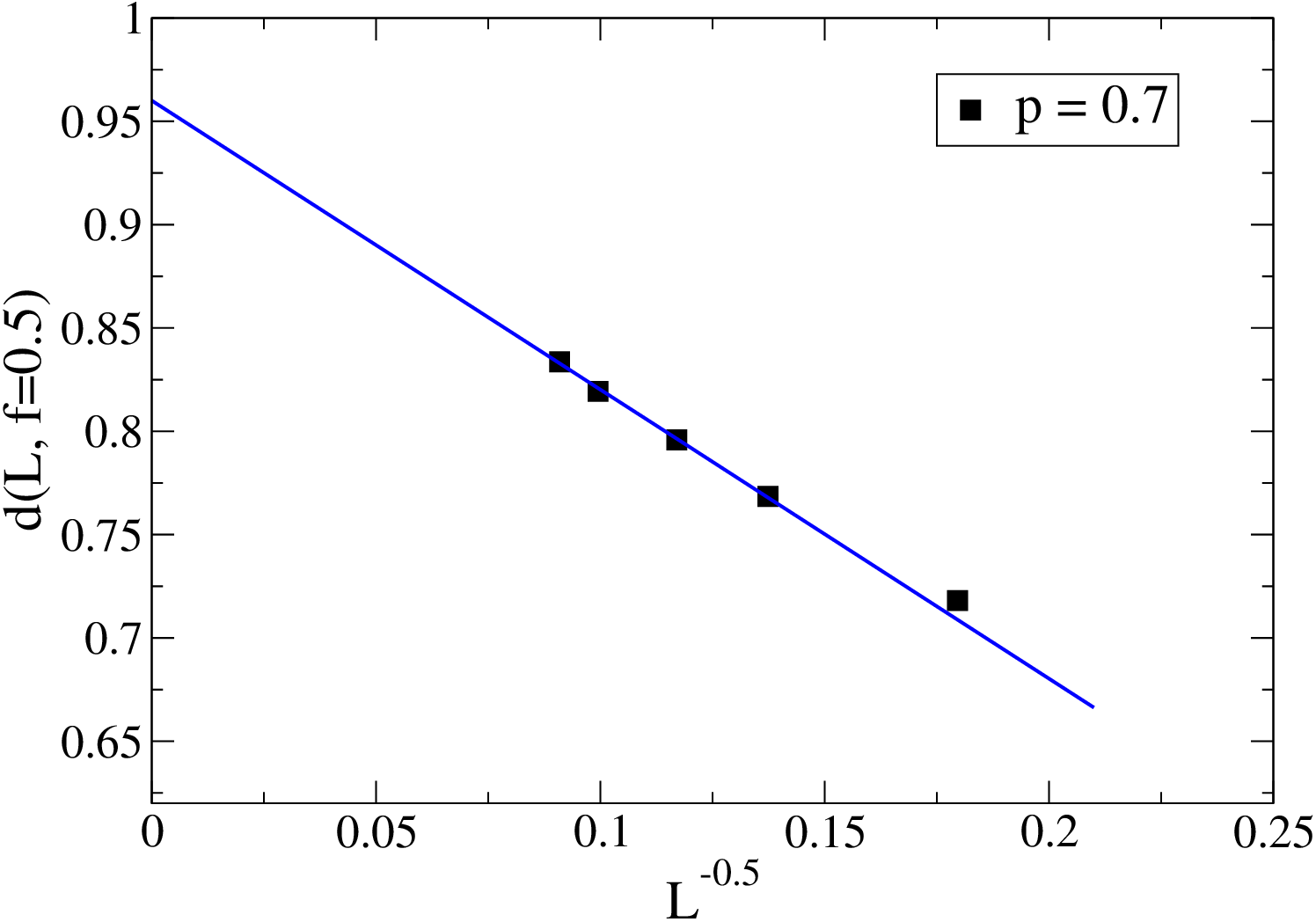}
\hspace{0.5cm}
\includegraphics[width=0.45\textwidth,angle=0]{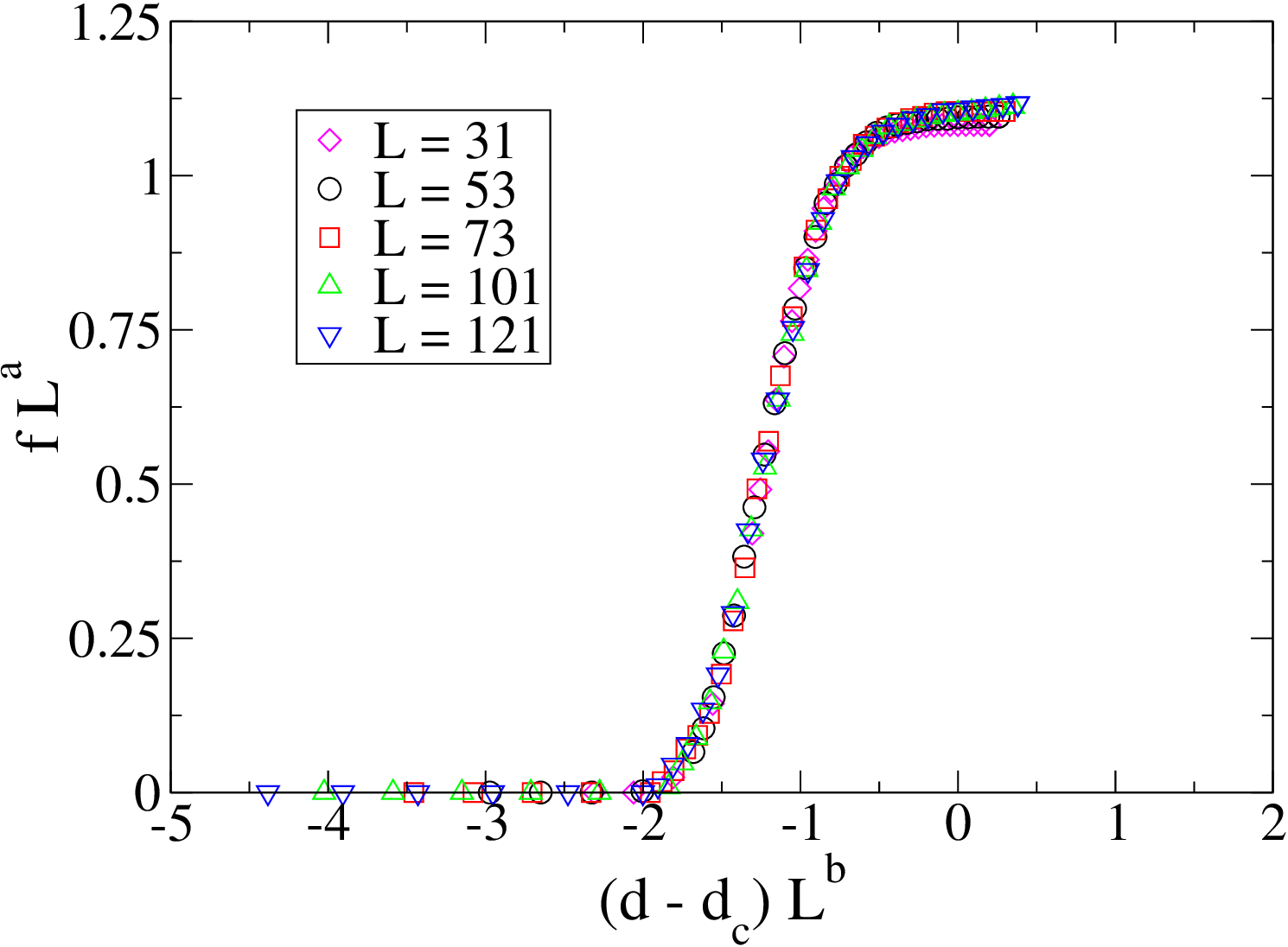}
\end{center}
\caption{The value of $d$ corresponding to the level $f=0.5$ for different sizes $L$, $d(L,f=0.5)$, as a function of $L^{-0.5}$ (left side). Fitting data, we obtained $d(L,f=0.5)=0.96007-1.3989\,L^{-0.5}$, which give us $d_{c}(p=0.7)\sim 0.96$ in the thermodynamic limit. It is also shown the scaling plot of $f$ (right side). The best collapse of data was obtained for $a=0.023\pm 0.002$, $b=0.47\pm 0.03$ and $d_{c}=0.960\pm 0.003$.}
\label{fig5}
\end{figure}

For the characterization of the phase transition, we have simulated the system for different lattice sizes $L$ and we have measured the fraction of samples $f$ which show all spins up when the initial density of up spins $d$ is varied in the range $0.4\leq d\leq 1.0$. In other words, this quantity $f$ gives us the probability that the population reaches consensus, for a given value of $d$. We have considered $1000$ samples for $L=31$ and $53$, $500$ samples for $L=73$ and $101$ and $200$ samples for $L=121$. The numerical results for $f$ as a function of $d$ for $L=101$ and typical values of $p$ are shown in Fig. \ref{fig3}. Observe that for values of $p$ near $0.6$ there is no transition, i.e., the consensus states are obtained only for $d=1$ in these cases. For the formal characterization of the transition, we can study how the density values $d$ corresponding to $f=0.5$ depends on the lattice size $L$. Taking the value $p=0.7$, for example, we exhibit in Fig. \ref{fig4} the results for $f$ as a function of $d$ for different lattice sizes $L$. We can estimate the values $d(L,f=0.5)$ from the crossing of the curves obtained from the simulations for each value of $L$ and the dashed straight line in Fig. \ref{fig4}. Plotting these values as a function of $L^{-b}$ one gets a straight line for $b\sim 0.5$ (see Fig. \ref{fig5}, left side). Fitting data, we obtained the equation

\begin{equation}\label{eq2}
d(L,f=0.5)=0.96007-1.3989\,L^{-0.5} ~,
\end{equation}
\noindent
which give us an estimate of the critical density $d_{c}\sim 0.96$ in the thermodynamic limit $L^{-1}\to 0$. The scaling plot of the quantity $f$ was obtained based on the equations
\begin{eqnarray} \label{fss1}
f(d,L) & = & L^{-a}\;\tilde{f}((d-d_{c})\;L^{b}) ~, \\ \label{fss2}
d_{c}(L) & = & d_{c}+c\;L^{-b} ~.
\end{eqnarray}
\noindent
The result is shown in Fig. \ref{fig5} (right side). The best collapse of data was obtained for $b=0.47 \pm 0.03$, $a=0.023 \pm 0.002$ and $d_{c}=0.96007\pm 0.0031$.

\begin{figure}[t]
\begin{center}
\vspace{1.0cm}
\includegraphics[width=0.5\textwidth,angle=0]{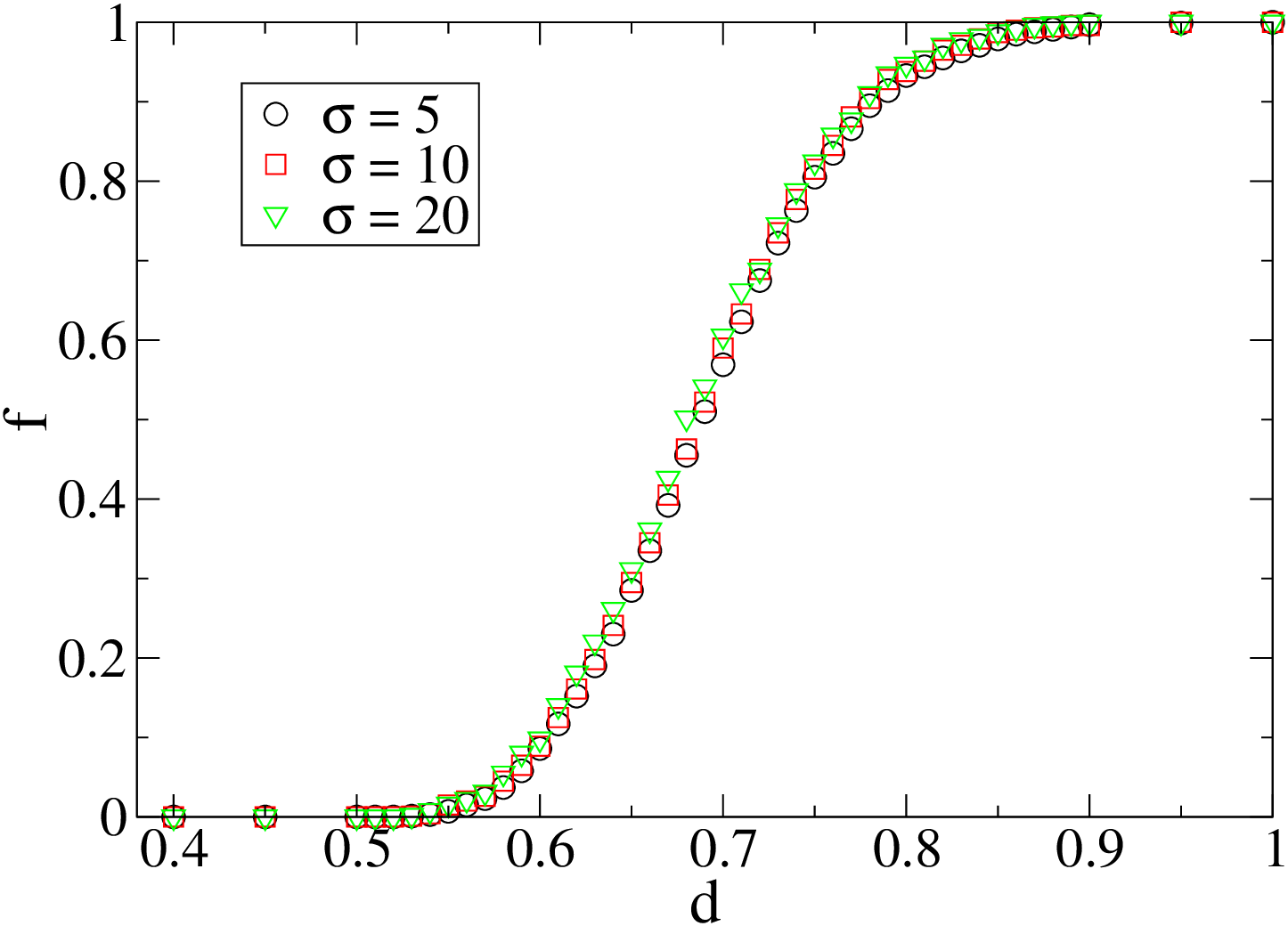}
\end{center}
\caption{Fraction $f$ of samples which show all spins up when the initial density $d$ is varied in the range $0.4\leq d\leq 1.0$, for $L=23$, $p=0.7$, $1000$ samples and typical values of the standard deviation $\sigma$ of the initial distribution of agents' reputation. This result shows that the increase of $\sigma$ does not change the behavior of $f$.}
\label{fig8}
\end{figure}

\begin{figure}[t]
\begin{center}
\vspace{1.0cm}
\includegraphics[width=0.5\textwidth,angle=0]{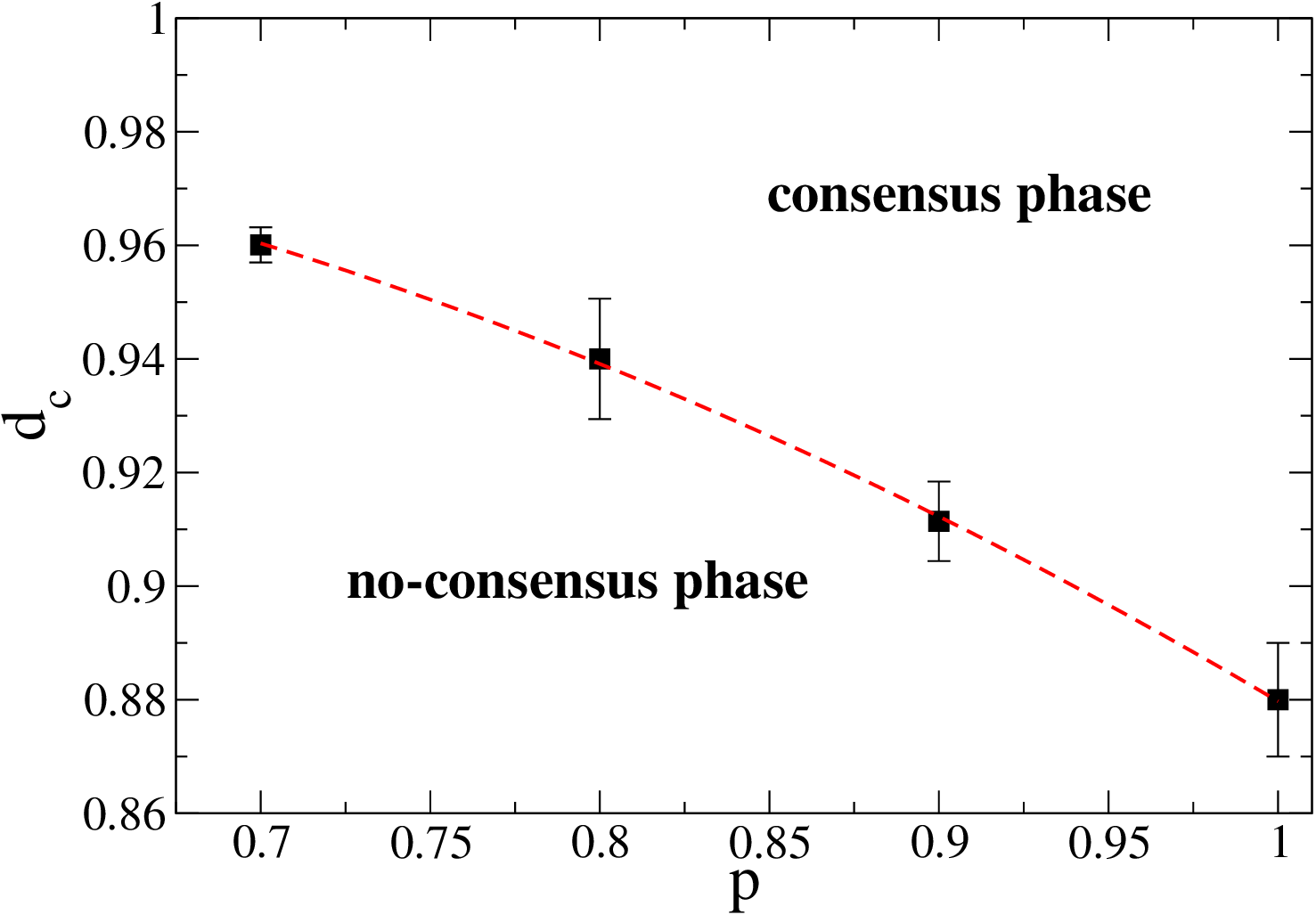}
\end{center}
\caption{Phase diagram of the model in the plane $d_{c}$ versus $p$. The points are numerical estimates of the critical densities $d_{c}$, whereas the dashed line is just a guide to the eyes. In the region above the curve the system always reaches consensus with all spins up, whereas in the region below the curve this kind of consensus never occurs.}
\label{fig9}
\end{figure}

Following the above-discussed process, we have estimated the critical points $d_{c}$ and the critical exponents $a$ and $b$ [see Eqs. (\ref{eq2}), (\ref{fss1}) and (\ref{fss2})] for other values of $p$. We have verified that the exponents $a$ and $b$ do not depend on $p$, considering the error bars. Thus, the criticallity of our model is defined by the critical exponents $a\sim 0.02$ and $b\sim 0.5$, and the phase transition occurs at specific critical points $d_{c}(p)$ which depends on the probability $p$. In addition, applying the process to identify the critical density $d_{c}$ [see Eq. (\ref{eq2})] for values smaller and near $p=0.7$, we have found that $d_{c}(p=0.69)=0.9845 \pm 0.0057$, but for $p=0.68$ we have found that $d_{c}(p=0.68)>1$, which suggests that for $p<0.69$ the system doesn't undergo the usual phase transition of the Sznajd model. This defines a critical value $p_{c} \sim 0.69$ below which the system does not present the phase transition between consensus and no-consensus states. Observe that this transition is robust with respect to the choice of different values of the standard deviation $\sigma$ of the initial distribution of agents' reputation (see Fig. \ref{fig8}). Taking into account the critical densities $d_{c}(p)$ calculated for typical values of $p$, we exhibit in Fig. \ref{fig9} the phase diagram of the model in the plane $d_{c}$ versus $p$. The points were estimated from the finite-size scaling analysis of the numerical results, whereas the line is just a guide to the eyes. Considering the thermodynamic limit $L\to\infty$, the critical points $d_{c}$ separate a region where the full-consensus states with all spins up always occurs, for each realization of the dynamics (above the frontier), from a region where this kind of consensus never occurs (below the curve).

\begin{figure}[t]
\begin{center}
\vspace{1.0cm}
\includegraphics[width=0.45\textwidth,angle=0]{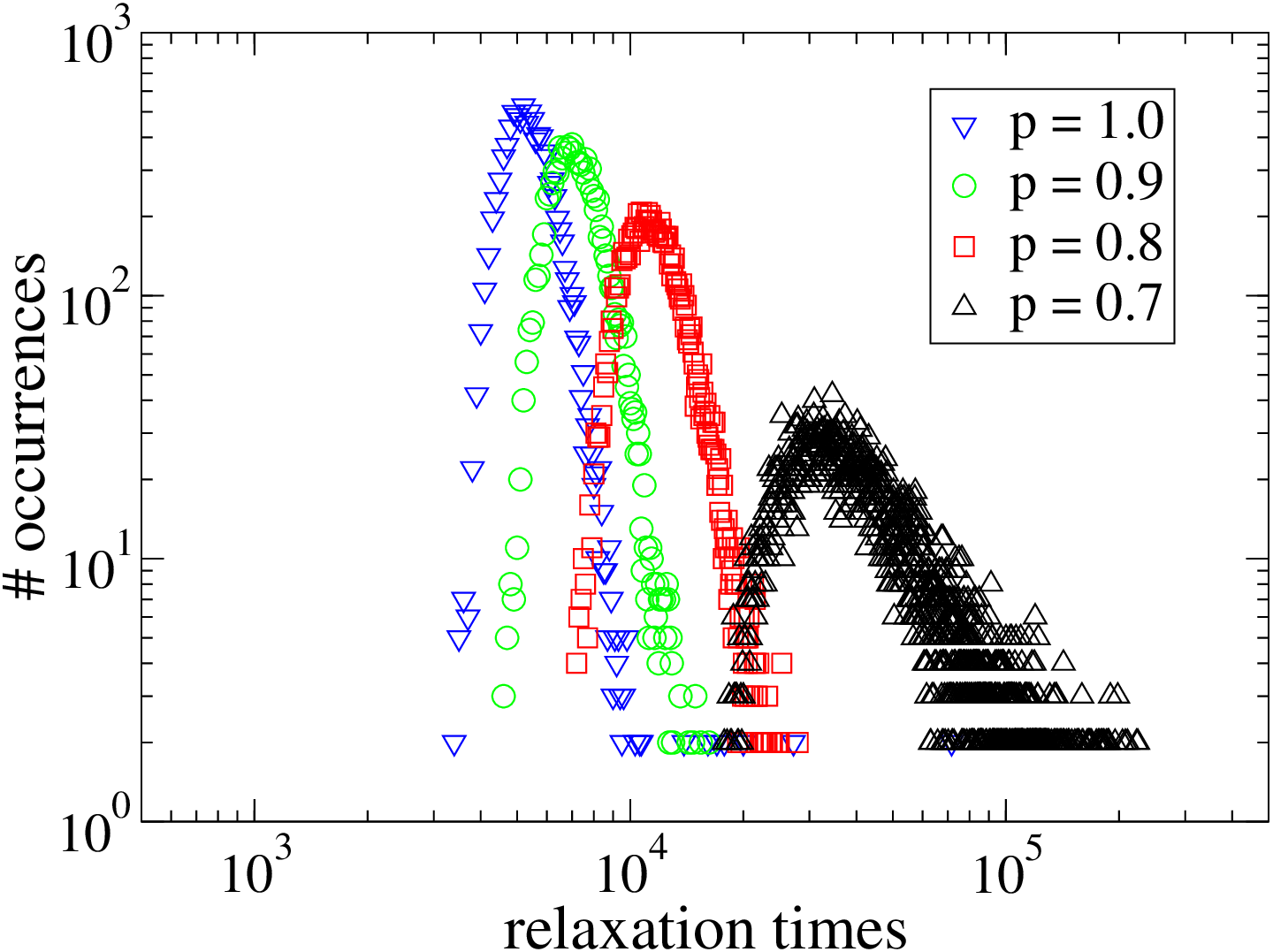}
\hspace{0.5cm}
\includegraphics[width=0.45\textwidth,angle=0]{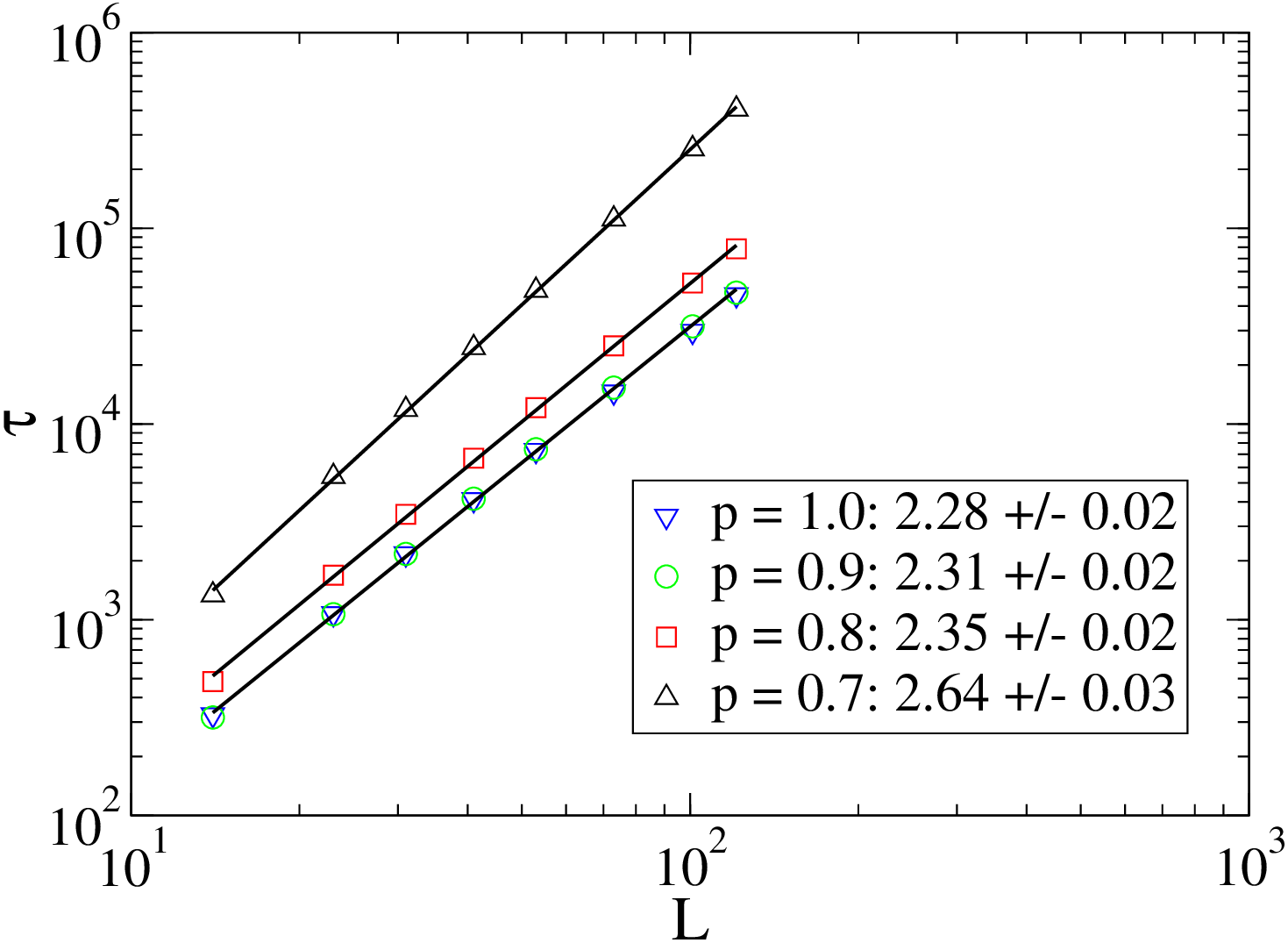}
\end{center}
\caption{Log-log plot of the histogram of relaxation times for $L=53$, $d=0.9$ and some values of $p$, obtained from $10^{4}$ samples (left side). The observed parabolas in the log-log plot indicate log-normal-like distributions for all values of the probability $p$. It is also shown the average relaxation times $\tau$ as functions of the lattice size $L$ (right side). Fitting data, we obtained the power-law behavior $\tau\sim L^{\alpha}$, where the numerical values of the exponents $\alpha$ are exhibited in the figure.}
\label{fig2}
\end{figure}

We have also studied the relaxation times of the model, i.e., the time needed to find all the agents at the end having the same opinion. The distribution of the number of sweeps through the lattice, averaged over $10^{4}$ samples, needed to reach the fixed point is shown in Fig. \ref{fig2} (left side), for typical values of the probability $p$. We can see that the relaxation time behavior is compatible with a log-normal distribution for all values of the probability $p$, which corresponds to a parabola in the log-log plot of Fig. \ref{fig2} (left side). In addition, the average relaxation times increase for decreasing values of $p$, as a consequence of the above-mentioned competition of reputations. It is also exhibited in Fig. \ref{fig2} (right side) the average relaxation time $\tau$ [also over $10^{4}$ samples, considering the relaxation times of Fig. \ref{fig2} (left side)] versus latice size $L$ in the log-log scale. We can verify a power-law relation between these quantities in the form $\tau\sim L^{\alpha}$, for large $L$, with a $p-$dependent exponent $\alpha$. Observe that $\alpha$ increases for decreasing values of $p$, which is also a consequence of the competition of reputations. Notice that log-normal distributions for the relaxation times and power-law relations between $\tau$ and $L$ were also found in previous works on the SM \cite{adriano,schneider,adriano2,sznajd_meu,schulze2}.


\section{Final Remarks}

\qquad In this work we have studied a modified version of the Sznajd sociophysics model. In particular, we considered reputation, a mechanism that limits the capacity of persuasion of the agents. In other words, a high-reputation group with a commom opinion may convince their neighbors with probability $p$, which induces an increase of the group's reputation. On the other hand, there is always a probability $q=1-p$ of the neighbors to keep their opinions, which induces a decrease of the group's reputation. These rules describe a competition between groups with high reputation and hesitant agents, which makes the full-consensus states (with all spins pointing in one direction) more difficult to be reached. The consequence is the occurrence of democracy-like situations, where the majority of spins point in one direction, for a wide range or parameters. 

We have observed a log-normal-like distribution of the relaxation time, i.e., the time needed to find all the agents eventually having the same opinion. In addition, the average relaxation times grow with the linear dimension of the lattice in the form $\tau\sim L^{\alpha}$, where $\alpha$ depends on the probability $p$. We have identified the critical density $d_{c}(p)$ for some values of $p$ by the finite-size scaling analysis of the fraction $f$ of samples which show all spins up when the initial density of up spins $d$ is varied. Our numerical results suggest that the system does not undergo the usual phase transition for $p<p_{c}\sim 0.69$.

It is important to notice that the inclusion of the new ingredient, i.e., reputation, makes the Sznajd model intrinsically dynamic. Whereas in the previous formulations of the SM the microscopic rules are static in a way that the dynamics is only related with the evolution of the system, in our case the SM becomes completely dynamic since the rules evolve with the system, i.e., the rules change according to the succession of visited states.

\vskip 2\baselineskip

{\large\bf Acknowledgments}

\vskip \baselineskip
\noindent
The authors acknowledge financial support from the Brazilian funding agency CNPq.

\vskip 2\baselineskip



\begin{thebibliography}{40}

\bibitem{axelrod} 
R. Axelrod, J. Conflict Resolut. \textbf{41}, 203 (1997).

\bibitem{baronchelli}
A. Baronchelli, M. Felice, V. Loreto, E. Caglioti, L. Steels, J. Stat. Mech P06014 (2006).

\bibitem{holley}
R. Holley, T. Liggett, Ann. Probab. \textbf{3}, 643 (1975).

\bibitem{galam}
S. Galam, Int. J. Mod. Phys. C \textbf{19}, 409 (2008).

\bibitem{sznajd_model}
K. Sznajd-Weron, J. Sznajd, Int. J. Mod. Phys. C \textbf{11}, 1157 (2000).

\bibitem{loreto_rmp}
C. Castellano, S. Fortunato, V. Loreto, Rev. Mod. Phys. \textbf{81}, 591 (2009).

\bibitem{pmco_book}
D. Stauffer, S. Moss de Oliveira, P. M. C. de Oliveira and J. S. S\'a Martins, \textit{Biology, Sociology, Geology by Computational Physicists} (Elsevier, Amsterdam, 2006).

\bibitem{stauffer_review}
D. Stauffer, J. Artif. Soc. Soc. Simul. 5 (2001) 1. Available at
http://jasss.soc.surrey.ac.uk/5/1/4.html

\bibitem{sznajd_review}
K. Sznajd-Weron, Acta Phys. Pol. B \textbf{36}, 2537 (2005).

\bibitem{adriano}
D. Stauffer, A. O. Sousa, S. Moss de Oliveira, Int. J. Mod. Phys. C \textbf{11}, 1239 (2000).

\bibitem{schneider}
J. J. Schneider, Int. J. Mod. Phys. C \textbf{15}, 659 (2004).

\bibitem{stauffer_2}
D. Stauffer, Int. J. Mod. Phys. C \textbf{13}, 315 (2002).

\bibitem{sznajd_meu}
N. Crokidakis, F. L. Forgerini, Phys. Lett. A \textbf{374}, 3380 (2010).

\bibitem{goldenfeld} N. Goldenfeld, C. Woese, Annu. Rev. Condens. Matter Phys. 2:375-99 (2011).

\bibitem{foot1} Notice that we will consider the integer part of the ratio $\sum_{i=1}^{4}R_{i}/4$.

\bibitem{edgardo} E. Brigatti, Phys. Rev. E \textbf{78}, 046108 (2008).

\bibitem{bonabeau} E. Bonabeau, G. Theraulaz, J. L. Deneubourg, Bulletin of Mathematical Biology \textbf{58}, 661 (1996).

\bibitem{bonabeau2} E. Bonabeau, G. Theraulaz, J. L. Deneubourg, Physica A \textbf{217}, 373 (1995).

\bibitem{stauffer_bonabeau} G. Weisbuch, D. Stauffer, Physica A \textbf{384}, 542 (2007).


\bibitem{foot2} In the following we will discuss the effect of changing $\sigma$ on the model.


\bibitem{adriano2}
A. O. Sousa, T. Yu-Song, M. Ausloos, Eur. Phys. J. B \textbf{66}, 115 (2008).

\bibitem{schulze2}
C. Schulze, Int. J. Mod. Phys. C \textbf{14}, 95 (2003).




\end{thebibliography}
\end{document}